# Software-based signal compression algorithm for ROM-stored electrical cables

**Tshimankinda Jerome Ngoy[1], Wa Nkongolo Mike Nkongolo[2],**
[1]Department of Electrical & Electronic, Faculty of Engineering, Independent Institute of Education (IIE) MSA, Johannesburg, South Africa
[2]Department of Informatics, Faculty of Engineering, Built Environment and Information Technology, University of Pretoria, South Africa

**ABSTRACT**

The purpose of this project is to determine a method for compressing the function codes located in the non-volatile memory of the on-board system after the linking/localisation phase, decompressing these functions, and executing them in the on-board volatile memory. The system and the execution of these functions are referred to as the uncompressed functions of the in-vehicle system. This approach ensures that the software is stored in ROM memory and can be restored after maintaining the runtime environment, protecting it from power surges or viruses occurring in the power cord. The decompression algorithm is stored in the ROM space through the main program, and a preventive method will be developed to compress the signal transmitted by the software over the cable. After the algorithm runs, the decompressed software is loaded into the ROM space to initialise the main program. By creating a backup copy, this method avoids the need to store software on an isolated server where viruses or power surges are less likely to occur. Since the main program initialisation software is stored in a compressed state in the ROM, and the decompression algorithm is compact, the project effectively utilises the ROM space. Additionally, excess energy present in the ROM can be harnessed by applying thermoelectricity (TE) to capture wasted thermal energy from the heated ROM, converting it into electrical energy to charge the battery.

***Corresponding Author:***

Tshimankinda Jerome Ngoy
Email: jerometshimak@gmail.com
Wa Nkongolo Mike Nkongolo
Department of Informatics, Faculty of Engineering, Built Environment and Information Technology
University of Pretoria
Hatfield 0028, South Africa
Email: mike.wankongolo@up.ac.za

## 1. INTRODUCTION

Currently, millions of kilometers of cables are being used to provide electrical connections in machinery, equipment, buildings and other places. If energy storage devices are used, they are completely separate from these power cords. However, if conductive software and energy storage can be integrated into the same cable, it will revolutionise energy storage applications [1]. Coaxial cable, is one of the most common and basic cable types used to transmit power or signals. Its internal conductor is surrounded by an electrically insulating layer and covered by an external tubular conductive shielding layer. Supercapacitors, also known as electrochemical capacitors, have become one of the most popular energy storage devices in recent years. Compared with other energy storage devices (such as batteries), supercapacitors have a faster charge and discharge rate, higher power density and longer service life [2]. In addition, a new *coaxial supercapacitor* (CSC) cable design has been demonstrated, which combines conduction and energy storage by modifying the copper core used for conduction.

In order to obtain the large area required for high supercapacitor performance, we developed a *normalized wavelet* (NW) on the outer surface of the copper wire. An interesting advantage of using the coaxial design is that electrical energy can be conducted through the internal conductive metal wire, and the electrical energy can be stored in the concentric layer of nanostructures added to the internal metal wire, and there is a layer of oxidation between the two. Therefore, the integration of cables and energy storage devices into a single unit provides a very promising opportunity to transmit power and store energy at the same time.

In one embodiment, this project includes a system for efficient use of Read-Only Memory (ROM) space in an integrated system. The project system includes an integrated system with a processor, and ROM. The decompression algorithm is stored in the ROM space through the main program. When the vehicle system is powered on, the decompression algorithm is executed by the processor. The decompression algorithm is suitable for the compression software stored in the ROM space [1].

The compression software includes the data needed to initialise the main program. After the algorithm runs, it loads the decompressed software into the ROM space to initialise the main program. Since the software for initialising the main program is stored in the ROM in a compressed state, and since the decompression algorithm is compact, the present invention effectively utilises the ROM space.

The space saved makes it possible to use a smaller and cheaper ROM. In addition, the space saved allows the use of a larger, more complex, and more feature-rich main program. In this case, the decompression must be separated from the execution of the decompression function, so that the decompression task can be handed over to another kernel, or it can be started as an execution earlier [1].

Any compression scheme should result in loss of information. This work should aim to quantify acceptable losses. This work must focus on the loss of frequency and amplitude information. Finally, it should be noted that compression only applies to relevant data. If the signal contains random noise superimposed on the fundamental wave, compression may be harmful. In addition, compression schemes used in the interference data domain of electrical systems should try to utilise the following knowledge: the input signal has very high energy at the fundamental frequency, and the waveform is highly periodic.

This project studies the design issues related to the realisation, adaptation and customisation of compression algorithms (especially data compression technologies aimed at increasing the energy of sensor arrays). The goal of this method is to reduce the consumption of non-volatile memory while keeping the CPU load low in critical parts. To use less non-volatile memory, uController [3] can be used. The whole process consists of two steps:

- Assuming decompression is performed at runtime, the existing in-vehicle system software can be updated to only support compression and decompression and as fast as possible.
- Apply the compression tool to the binary/hexadecimal files obtained after the linking/localisation stage.

Specific theoretical examples of Infineon Tricore uControllers used and the amount of FLASH memory available are [3]:

- TC1734 - 1MB FLASH,
- TC1738 - 1.5MB FLASH,
- TC1767 - 2MB FLASH

After all optimisations, if the flash memory capacity required by the software project is 1.1MB, the company will choose TC1738 uController [3]. But by using the proposed method (compressing multiple functions), the required FLASH size can be reduced to less than 1MB, so that the cheaper TC1734 can be used. The gain is multiplied by the number of embedded systems produced using uController. In fact, for software projects and microcontroller series with different flash memory sizes, it is worthwhile to compress to check whether the same series of lower-level uControllers (with less flash memory) can be used after compression.

Due to decompression, we do not want to increase the *central processing unit* (CPU) load during the execution of the key code, so it is mainly suitable for functions that are executed only once (configuration, initialisation, and control functions). There are also no restrictions on the applicability of repeated items. In Figure 1, the setup is displayed. In this experiment, thermoelectric generator (TEG) was installed between the hybrid heat sink cooler and the CPU processor to produce micro-scale electricity. In our experiment, we employed an Intel Core i3-2100 processor, which when forced to operate using *overclock checking tool* (OCCT) software may produce heat up to 80 $^{o}$C.

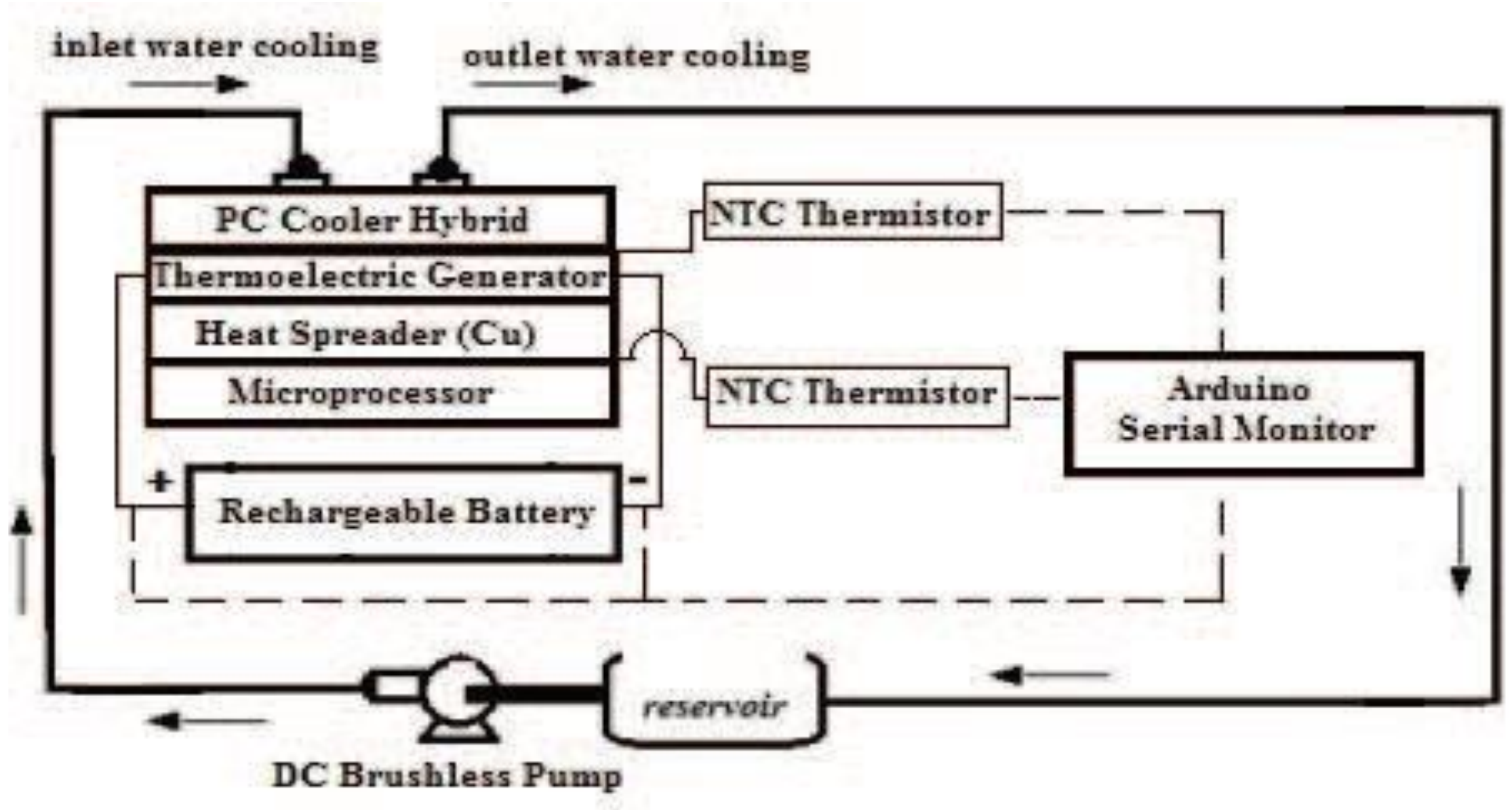

Figure 1. Schematic configuration of TEG on CPU processor [7].

The executed function or external event triggers the executed function, but this is only when the CPU load introduced by the decompression mechanism does not affect the system function. There is a restriction on the functions we want to compress: the binary code obtained for these functions must be relocatable because the execution address is different from the build address. Generally, the binary result used for relative addressing will take up less memory space [4]. The general steps for compressing and decompressing data are shown in Figure 2.

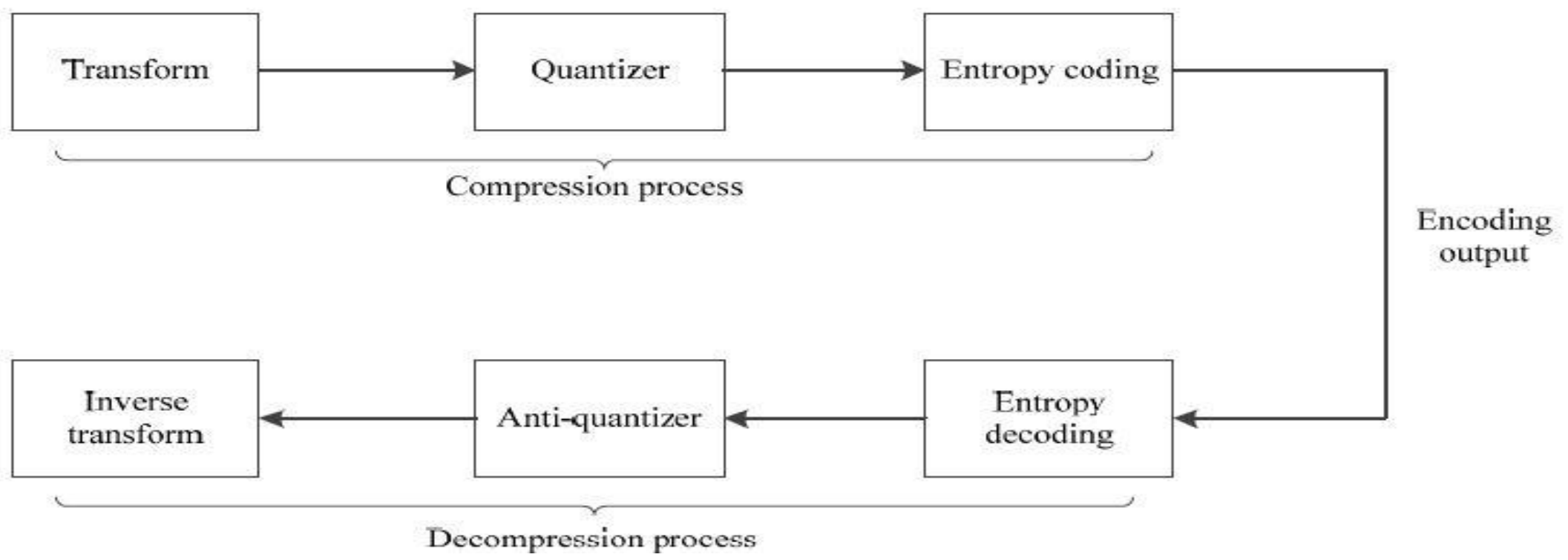

Figure 2. Block diagram of data compression and decompression.

Obviously, a compression algorithm with a certain degree of complexity is worth exploring [17, 18, 19]. Also note that these calculations only consider the local power consumption on the compression node. Downstream energy saving can further share the compressed time/energy expenditure [5]. The current and future memory size limitations of sensor processors require reconsideration of memory usage in compression calculations. Although each generation of sensor processor technology shows higher capacity, they usually still provide less than 50KB of code storage space and less data RAM. Therefore, the compression algorithm originally used on desktops or servers must be redesigned to reduce confusion, code size and dynamic memory usage.

This project evaluates a method and system to reduce the amount of ROM required by digital embedded systems, and explores methods of lossless compression without loss because it is suitable for a wider range of applications, datasets, and user. As shown in Figure 3, the principal block diagram of the software for electrical cables stored in ROM. First, at the substation the current of the step-down AC-DC or DC-AC voltage regulator is processed by demodulator or modulator signal of the machine which is separated from background noise. Then, the signal is compressed using an algorithm to process the signal in the computer or CPU cooler and later stored in ROMs.

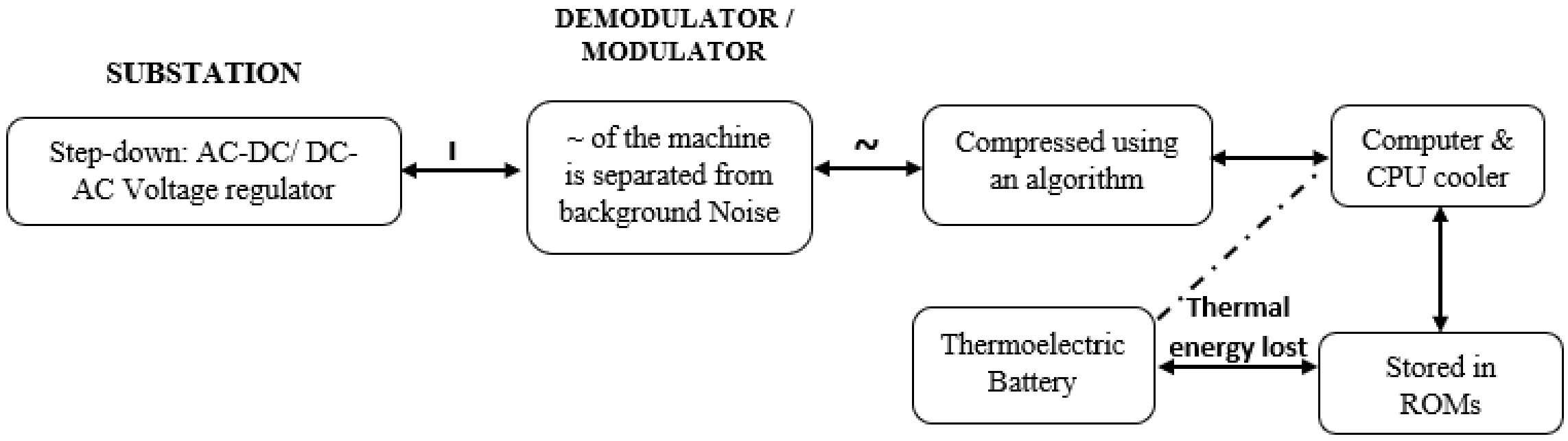

Figure 3. The search function framework.

Data compression covers technologies that can represent information in a compact form. These compact representations are obtained by identifying and using the structure that exists in the data. When digitising a constant envelope sine wave, we will spend a lot of bits to encode its samples. However, we can represent this signal in a compact form in terms of amplitude, frequency, and phase.

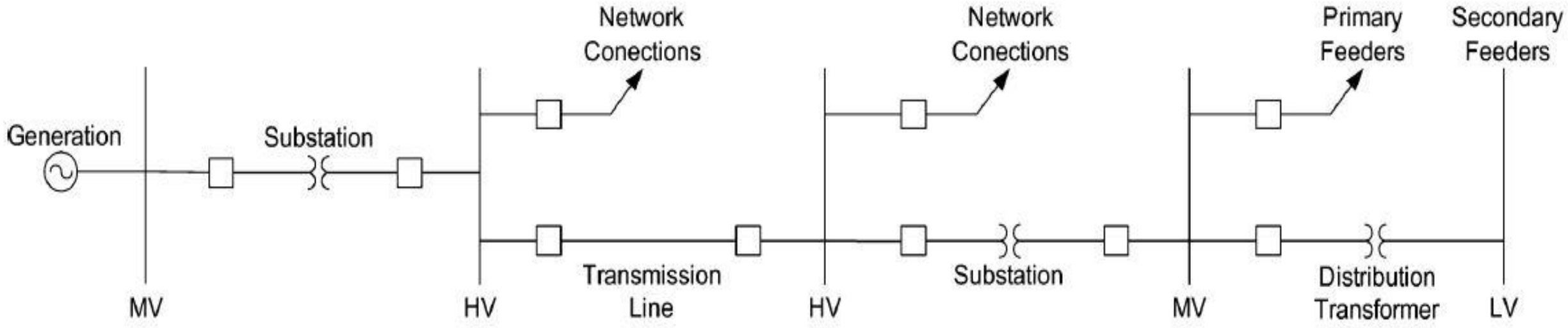

Figure 4. A typical power system scheme (LV—low voltage, MV—medium voltage, and HV—high voltage) [15].

A simplified diagram of the power supply system is shown in Figure 4. Usually, the generator is powered at the level of 13.8 kV, and the voltage is increased in the power station, so the energy is transmitted up to 1000 kV using high-voltage transmission lines in the range of 138, which are called *high-voltage* (HV) and *ultra-high-voltage* (UHV), respectively [6]. When the energy reaches the distribution station, the voltage drops again to the *medium voltage* (MV) level, which is a characteristic of the distribution network in the system. Generally, the level of the primary distribution line starting from the substation is 13.8 kV (MT) and the length is less than 10 km, but it may be longer in rural areas because the power demand is relatively scarce and scattered. The distribution transformer is connected to the primary feeder at many points to reduce the voltage level from 13.8 kV to 127 V, 220 V or 380 V (approximately) to supply power to the end user. Then, the secondary power distribution system corresponds to the *low voltage* (LV) feeder.

## 2. RESEARCH METHOD

The idea is not to focus on the compression method (it can be other innovative methods), but on how to implement all the stages described in the embedded project in a simple and fast way. Generally, data can be compressed by eliminating data redundancy and irrelevance. Modeling and coding are two levels of compressed data: at the first level, the data will be analysed to obtain any redundant information, and it will be extracted to develop the model. In the second level, the difference between the modeled data and the actual data (called residual) is calculated and encoded by coding techniques.

For compression, the ZLLIB library was used [7], which it is a free, lossless data-compression library which can be used on almost any computer hardware and operating system. After studying several embedded software systems, we found that most functions can be relocated, or the compiler may be forced to generate relocatable code. Generally, the binary result used for relative addressing will take up less memory space [4]. If the algorithm runs on another powerful code execution core of the same embedded system (such as PCP or auxiliary core on Infineon Tricore uControllers), the decompression of this function can be performed in parallel with the execution of other functions. uController starts earlier than execution when it is idle. Taking these facts as input, the decompression function must be separated from the execution of the decompression function in order to pass the decompression task to another kernel or start it earlier in execution.

Both methods are designed to accelerate decompression. In order to evaluate the efficiency of profile-driven and differential compression schemes presented in the previous sections, we compare their performance to those of some of the most known compressors. In particular, we chose two variants of Lempel and Ziv text replacement encoders [8]: the LZSS algorithm of Storer and Szimansky [9] and the LZAR method that combines LZSS with arithmetic coding [10]. LZSS is a byte-oriented compressor, which assumes a ring buffer, which initially contains zero bytes. It reads multiple bytes from the buffer, if convenient, find the longest string corresponding to the last read byte in the buffer, and use the corresponding length and position in the buffer to encode it in binary. The length of the unencoded string LZAR improves on LZSS by taking advantage of the fact that not all bytes have the same frequency of appearance in a cache line. Therefore, higher frequency bytes are encoded with fewer bits, and lower frequency bytes are encoded with more bits; therefore, the total length of the cache line being compressed will be reduced.

The arithmetic code is used to generate a variable length pattern that encodes the length and position of compressed bytes. We note that for two LZ type compressors, slots of size S = 8 and S = 12 can be used in the compressed storage area. The results for the case of S = 12 are reported in Section 4, while the data for S = 8 is omitted due to space constraints.

The hardware-assisted compression scheme of this research is implemented in the Simplecalar simulation framework [11]. In particular, we adopted the functional cache simulator sim-cache as the simulation engine. We consider a system with compressed main memory (that is, the information is stored in the cache in an uncompressed format), where the compression hardware is located between the cache and the main memory. Here, the fundamental difference between code compression and data compression is that for the latter type of compression, decompression is required while the program is running, while for the former type of compression, only decompression is required. This fact has a profound effect on applicable compression algorithms and architectures: for example, it excludes highly asymmetric schemes in which compression is more involved than decompression. The potential for converting heat into electrical energy in computers has not been extensively studied. In this research, we implemented a thermoelectric generator module, which is combined with the hybrid cooling system on the CPU processor to convert heat into electrical energy, which is the main key part of the computer system, it generates a lot of heat, and its work is closely related to temperature. In addition, thermoelectricity has also been applied to other fields, such as biomass heating [12] and processors in CPU [13]. In our research, we explored the potential development of using thermoelectric generators to generate a new generation of electrical energy from microprocessor-based computers. We observe 0.5W output generated by a single thermoelectric module. Figures 6 and 7 show the *Lempel-Ziv-Welch* (LZW) implementation of our S-LZW sensor node, the number of instructions and the default main memory usage of the algorithm evaluated in [14].

*Lempel–Ziv–Oberhumer* (LZO) is an exception. In [14] they said that of amongst all the algorithms they evaluated, LZO has the lowest power consumption for compressing and sending 1MB of data. The developers of LZO have implemented a version specifically for embedded systems (for example, the version we call miniLZO). It should be noted that in the configuration of the system with the processing according to this embodiment, the sequence length encoding algorithm is used to compress the initvars and Zerovars parts, obtaining an average compression ratio of 10:1 and saving 90 KB of ROM space in the target system. The decompression algorithm consists of 24 ARM6 instructions, so it occupies less than 100 bytes of code in the ROM of the target system. In this embodiment, JumpStart 2.2TM is used as a development kit. A better compression algorithm (such as LZWTM) can achieve a compression ratio of 20:1, but the cost is that the code in the ROM space of the target system exceeds 5 KB.

Therefore, there is a need for a solution to eliminate the wasted address space of the on-board system. The required solution should reduce the amount of ROM space required to store runtime environment information. The required solution is to use ROM space more efficiently, reduce the cost of the onboard system, and provide greater functionality for the main program code. The present invention provides this required solution. In order to solve the problem of periodic signal compression, we propose an effective lossless compression method for periodic signals based on an adaptive dictionary. When a periodic signal is predicted, the historical information of the signal is important. We have built a dictionary to store signal history information because the historical information of the signal is rich. The data gathered thus far for the compression of electrical signals is summarised in Table 1. We maintained the evaluation measures used by the various strategies in Table 1. As can be seen, deciding which metric will be used to assess the various lossy compression algorithms is a crucial challenge for the development of approaches for compressing electric signals.

Table 1. Comparison of some techniques used to compressed electric signals [15].

| Group | Category | Basic Technique | Compression Ratio | Distortion Metric | Distortion Value |
|---|---|---|---|---|---|
| Lossless | 1D | Lempel-Ziv | 5:1 | – | – |
| | | Delta-modulation Huffman Coding | 2:1 | – | – |
| | 2D | JPEG2000 | 9:1 | – | – |
| Lossy | Wavelet Transform | Daubechies DWT | 6:1 to 3:1 | NMSE | $10^{-5}$ to $10^{-6}$ |
| | | Daubechies DWT | 3.43:1 | NMSE | $10^{-4}$ |
| | | Slantlet DWT | 10:1 | MSE | -19 dB |
| | | B-spline DWT | 15:1 | MSE | -25 dB |
| | Wavelet Packet | WPT and LZW | 10:1 | PRD | 10% |
| | | WPT and Arithmetic Coding | 6.9:1 | NMSE | $10^{-5}$ |
| | | EZW | 10:1 to 16:1 | NMSE | $10^{-5}$ |
| | Mixed Transform and Parametric | Fundamental, Harmonic and Transient Coding | 16:1 | MSE | -30 dB |
| | Parametric Coding | Damped Sinusoids Modeling | > 16:1 | SNR | > 31 dB |

## 3. RESULTS AND DISCUSSION

### 3.1. Implementing Thermoelectric Generator on CPU

The experiment is to observe the voltage generated when TEG is installed in the processor CPU. To prevent overheating, TEG has been connected to a hybrid radiator cooler, which consists of an aluminum radiator with water flowing in the radiator tube. We used Epcos NTC thermistor model B57891 to detect the temperature of the processor and TEG. All measurements are monitored and recorded using Arduino Uno. Figure 5 displays the current and voltage produced by the CPU processor. The maximum temperature attained is 53 ° C, and the maximum current and voltage are 190 mA and 2.4 V, respectively.

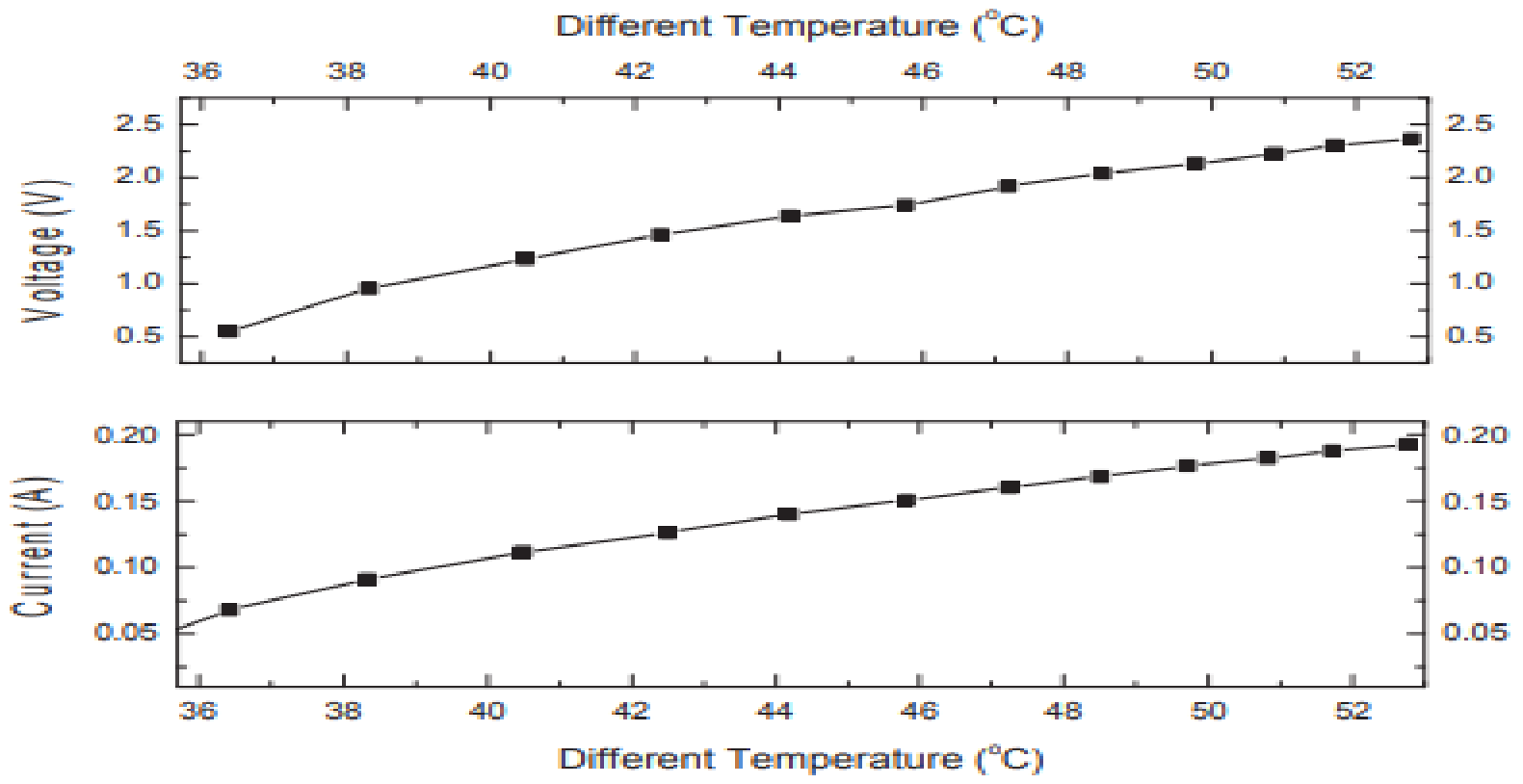

Figure 5. (a) Current output and (b) the thermoelectric generator's voltage as implemented by the CPU processor.

The excess energy present in the ROM can be utilised by applying thermoelectricity (TE) to capture the lost electrons (wasted thermal energy) from the heated ROM. The recovered thermal energy is then optimised by being converted into electrical energy for charging the battery.

### 3.2. Signal compression by LZW

Figure 6 verifies the efficiency and adaptability of the method proposed in this project.

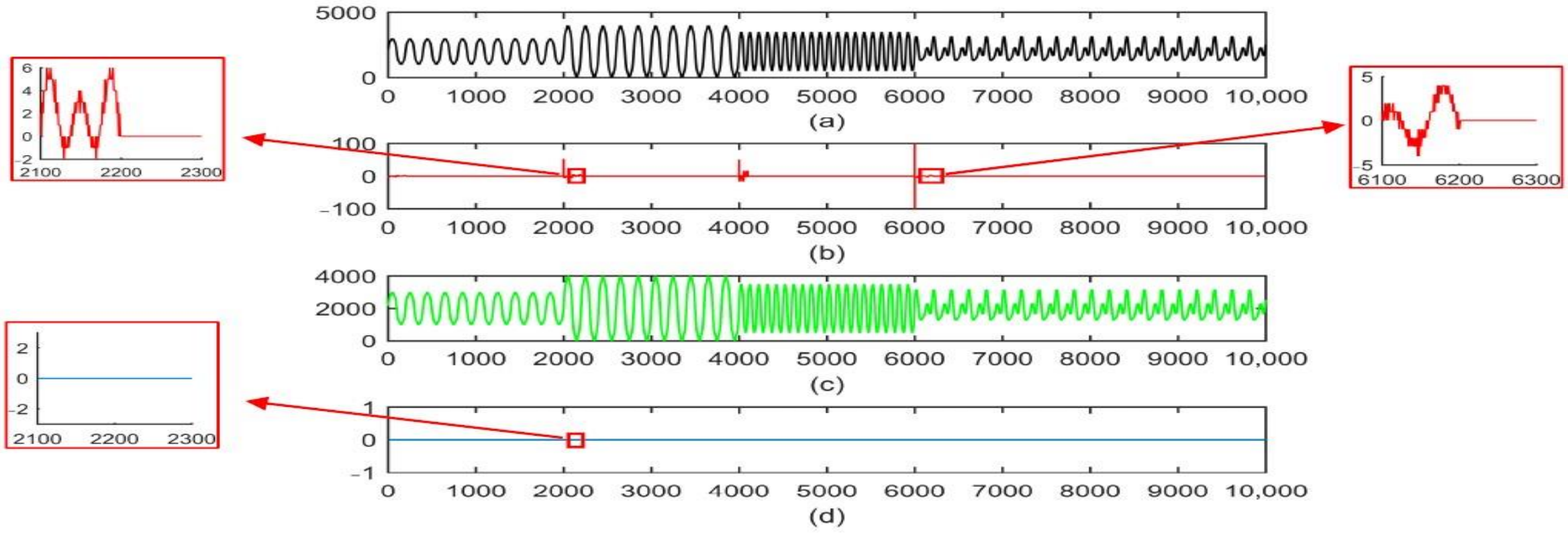

Figure 6. (a) Original signal (b) Coding output (c) Decoding output (d) Error between the decoding output and the original signal.

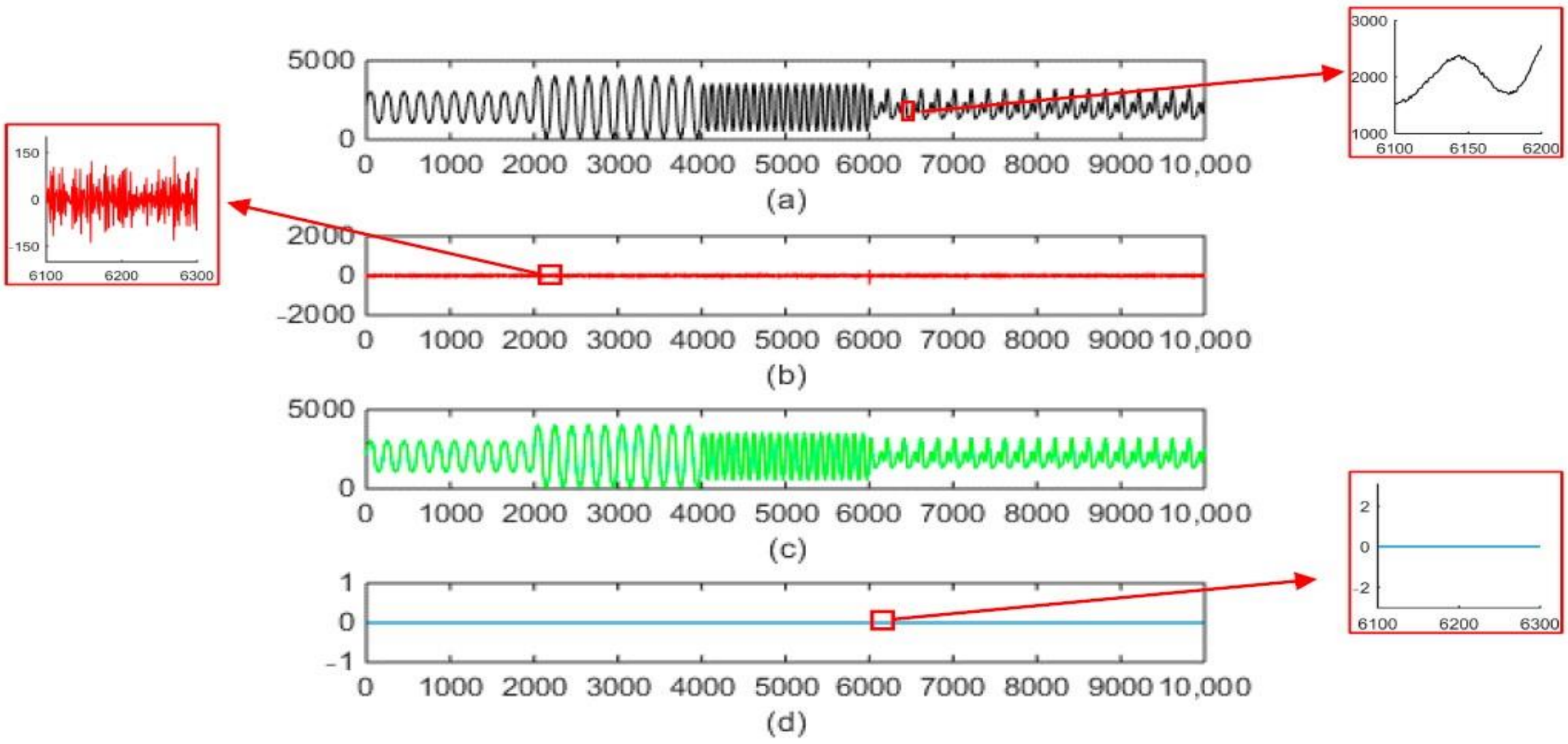

Figure 7. (a) The signals containing additive noise (b) Coding output (c) Decoding output (d) Error between the decoding output result and the original signal.

Then, noise is added to the original signal in Figure 6, and its signal-to-noise ratio is 34.2 dB. In Figure 7, the prediction effect of the algorithm on the periodic signal with added noise is shown, and the prediction effect of the noise is worse than that of Figure 7. In fact, the method proposed in this article is a lossless compression algorithm [16]. We have added noise to the periodic signal, and the signal amplitude is random within a certain range. It is difficult for the prediction model to use context information to accurately predict the amplitude of the signal, so the output prediction residual will fluctuate in a small range. Therefore, the signal in Figure 6 is also the next depth search direction of the algorithm.

We can see that the coding effect of this method on periodic signals in this study is far superior to the other three methods. When the encoding output of the method proposed in this paper is compressed by LZW, the compression rate is much higher than the other three methods. Moreover, the method provided in this article belongs to lossless compression. When compressing the signal by this method, the signal information will not be lost. From the decomposition, we notice how the energy of memory access dominates the total energy budget. On the contrary, the cost of the compressor is almost negligible. This tells us that if the main goal is energy optimisation, it may be interesting to study hardware implementations of more complex (and more efficient compression) schemes (such as the LZ-type method we discuss in this article). In fact, the additional complexity and energy consumption of the compression unit will be offset by the savings generated by reducing memory traffic.

### 3.3. Main compression techniques for electric signals

A voltage dip signal that was generated by employing the *Discrete Wavelet Transform* (DWT) with Daubechies four coefficient filters is shown in Figure 8 together with the detail bands and the approximation band of the coarsest scale. Be aware that the signal's many occurrences, particularly transients, can be captured by the wavelet transform. Sinusoidal signals are thus inappropriate for an effective representation in the wavelet transform domain due to their small bandwidth and variable frequency. On the other hand, as seen in Figure 8, it has the capacity to catch transient components. This served as the impetus for the study of hybrid coding methods.

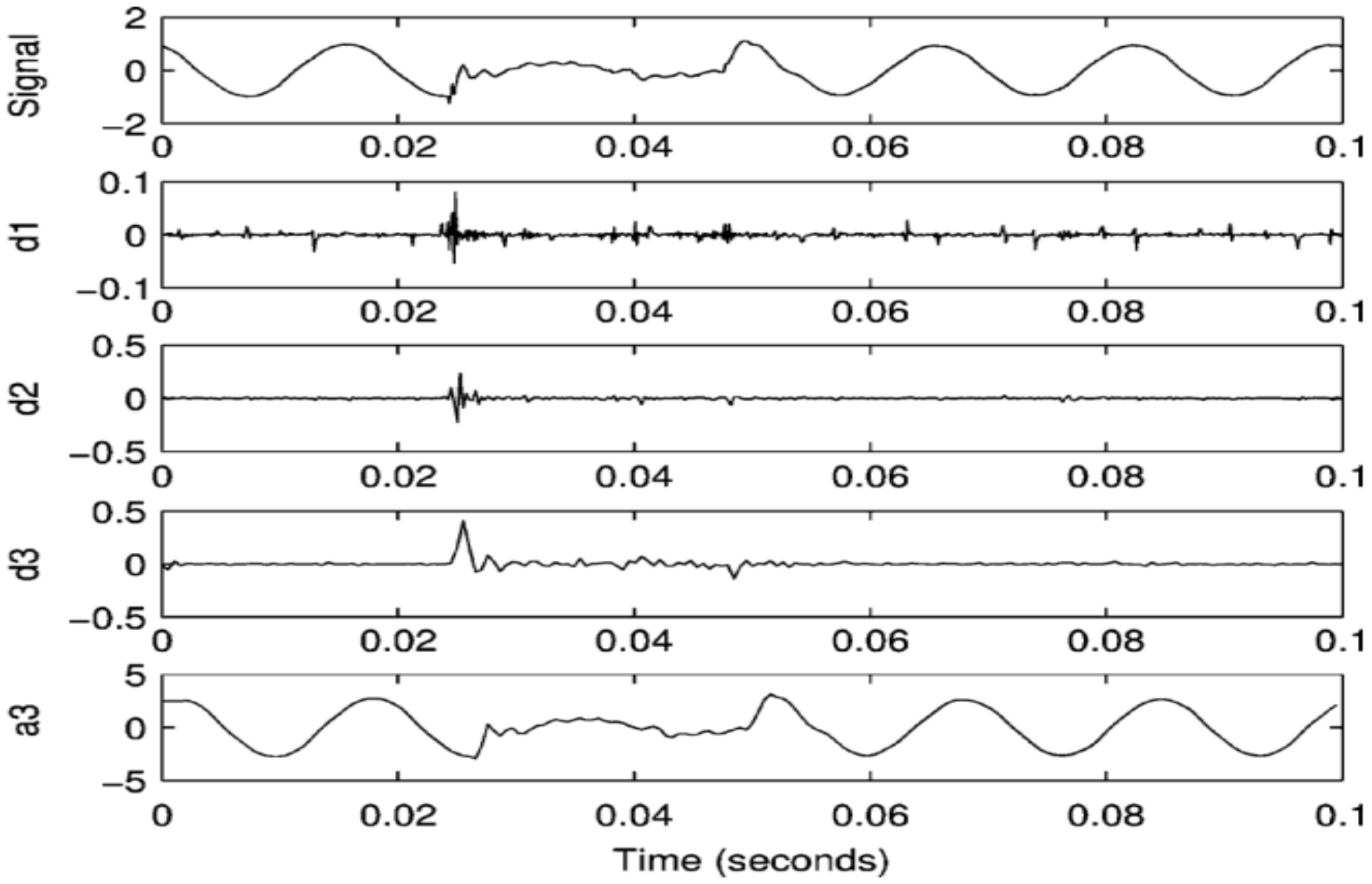

Figure 8. Detail and approximation bands for an IEEE project group 1159.2 voltage dip signal with a sampling rate of 15 360 Hz. The original signal is represented by the top plot, and the detail bands are displayed from top to bottom in decreasing frequency (increasing scale) order. The bottom plot corresponds to the coarsest scale's approximation band.

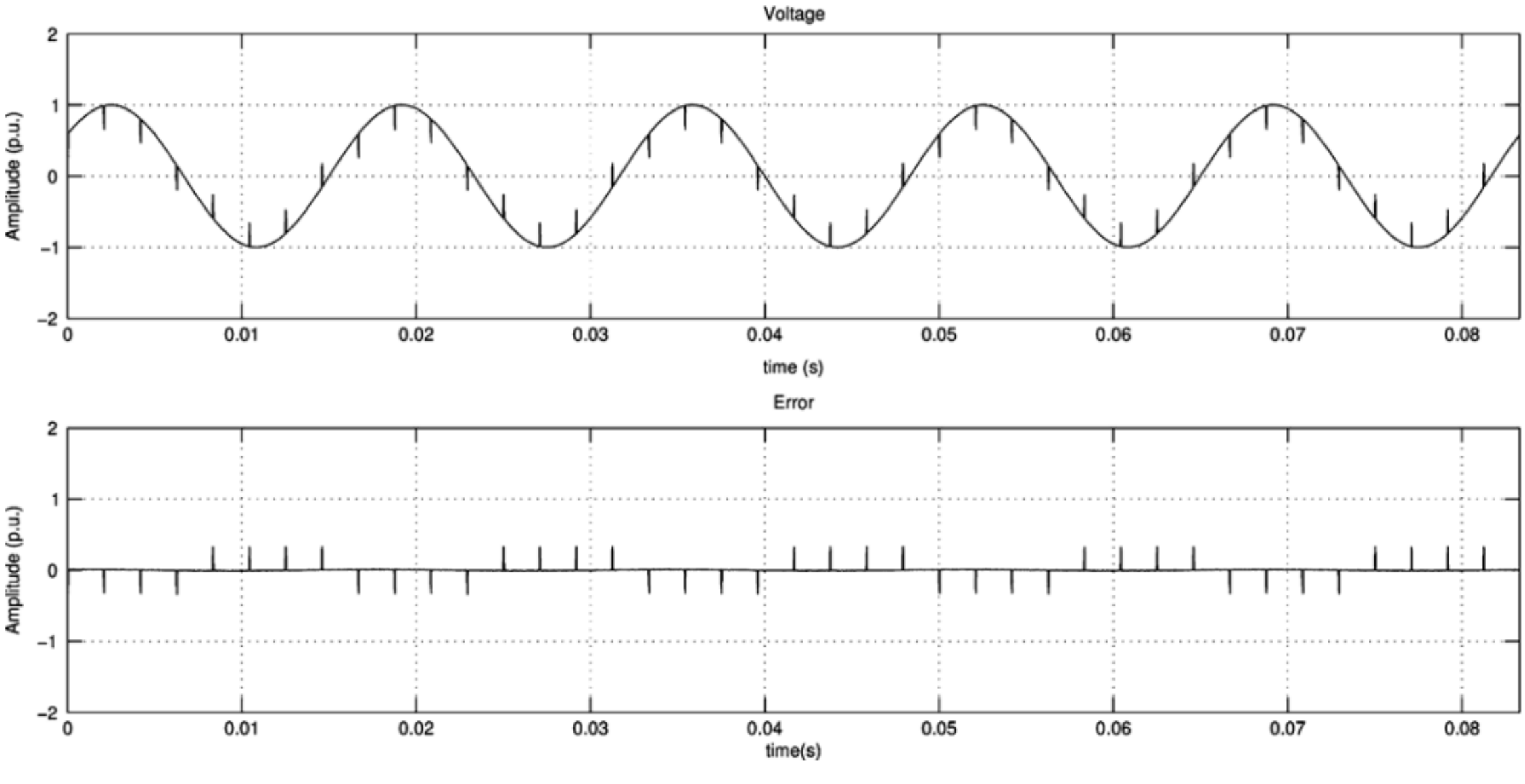

Figure 9. Voltage signal that has been tainted by recurring transient events and residue that is generated by removing the sinusoidal component.

Figure 9, which depicts a voltage signal and the residue left over after subtracting the fundamental component, serves as an illustration of this. Improved performance can be achieved since the fundamental component—the sinusoidal one—can be fully set using five parameters (beginning and ending samples, amplitude,

frequency, and phase) [15]. Figure 10, shows a flow chart of the steps of a process in accordance with one embodiment of the project.

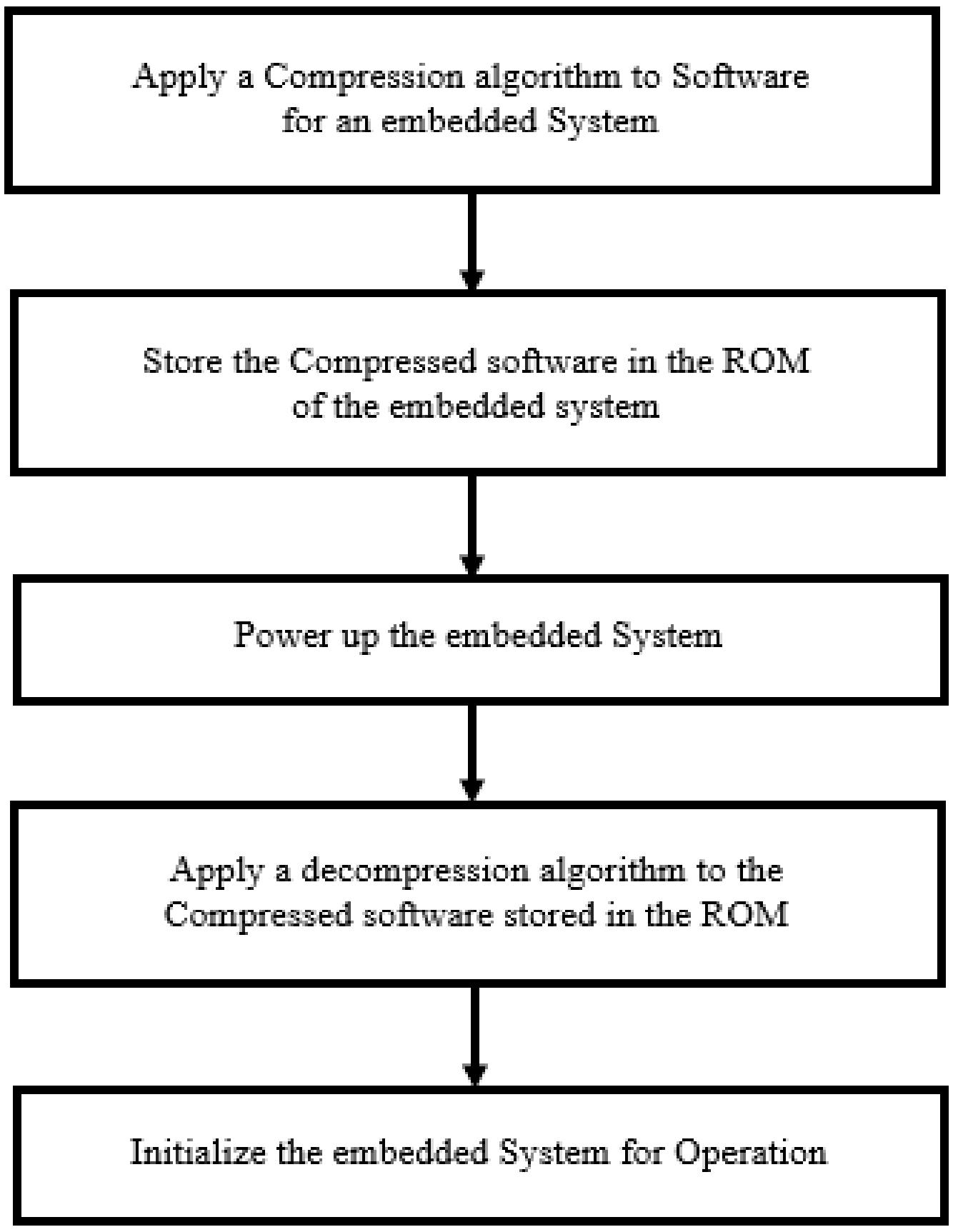

Figure 10. A flowchart showing the steps of a process according to one embodiment.

## 4. CONCLUSION

This project addresses the issue of periodic signal compression by proposing a novel adaptive coding method. The encoded output is compressed using the LZW algorithm. To minimise energy consumption in embedded kernel-based systems, we introduce hardware-assisted data compression. Our approach involves a new architecture model specifically designed for data compression, along with various compression methods that are well-suited for this model. By implementing the selected compression scheme, we observe a reduction in memory communication volume and power consumption in the cache memory path of the system running the standard benchmark. The achieved reduction ranges between 4.2% and 35.2%. This strategy aims to facilitate the adoption of smart sensing, monitoring, measuring, diagnostics, and protection in new low-cost device types, thereby enhancing electrical signal compression technology and improving the grid. Additionally, we explore the utilisation of excess energy present in the ROM. We propose harnessing thermoelectricity (TE) to capture wasted thermal energy in the heated ROM, converting it into electrical energy. This recovered thermal energy is then optimised for charging the supercapacitor, enabling efficient energy utilisation..

## ACKNOWLEDGEMENTS

The authors express their gratitudes to the University of Pretoria's Faculty of Engineering, Built Environment, and Information Technology for their support in funding this research project through the Doctorate UCDP Grant A1F637.

## REFERENCES

[1] S. Ott, "Method and system for using decompression on compressed software stored in non-volatile memory of an embedded computer system to yield decompressed software including initialized variables for a runtime environment." U.S. Patent 6,023,761, issued February 8, 2000.D. Jovcic, “Series LC DC circuit breaker,” *High Volt.*, vol. 4, no. 2, pp. 130–137, Jun. 2019, doi: 10.1049/hve.2019.0003.

[2] a) M. Winter, R. J. Brodd, Chem. Rev. 2004, 104, 4245; b) J. R. Miller, P. Simon, Science 2008, 321, 651; c) J. R. Miller, A. F. Burke, Electrochem. Soc. Interface 2008, 17, 53; d) M. D. Stoller, S. Park, Y. Zhu, J. An, R. S. Ruoff, Nano Lett. 2008, 8, 3498; e)

L.-Q. Mai, F. Yang, Y.-L. Zhao, X. Xu, L. Xu, Y.-Z. Luo, Nat. Commun. 2011, 2, 381; f) B. Duong, Z. Yu, P. Gangopadhyay, S. Seraphin, N. Peyghambarian, J. Thomas, Adv. Mater. Interfaces 2014, 1, 1300014.
[3] Infineon TC I 967, TCI734, TCI738 3 2-bit Single-Chip uController user manuals..
[4] K. Saito, M. Kamei, H. Ishima, "Execution program generation method, execution program generation apparatus, execution program execution method, and computer readable storage medium", US Patent US6715142BI, Mar. 3 0, 2004.
[5] J.K. Singh. "Method of preventing software piracy by uniquely identifying the specific magnetic storage device the software is stored on." U.S. Patent No. 5,615,061. 25 Mar. 1997
[6] M.P. Tcheou, L. Lovisolo, M.V. Ribeiro, and E.A. Da Silva, M.A. Rodrigues, J.M. Romano, P.S. Diniz. "The compression of electric signal waveforms for smart grids: State of the art and future trends". IEEE Transactions on Smart Grid, vol. 5, no. 1, pp. 290-302, 2013.
[7] N.H. Pranita, K. Azura, A. Ismardi, and T.A. Ajiwiguna, I.P. Handayani. "Implementing thermoelectric generator on CPU processor". In2015 International Conference on Control, Electronics, Renewable Energy and Communications (ICCEREC) IEEE, pp. 108-111. 2015.
[8] J.L. Gamy (compression) and M. Adler (decompression). In ZLLIB library. http://zlib. Net/
[9] J. Ziv, A. Lempel, \A Universal Algorithm for Sequential Data Compression," IEEE Trans. on Information Theory, Vol. 23, No. 3, pp. 337-343, 1977.
[10] J. A. Storer, Data Compression: Methods and Theory, Computer Science Press, 1988.
[11] I. Witten, R. Neal, J. Cleary, \Arithmetic Coding for Data Compression", Comm. of the ACM, Vol. 30, No. 6, pp. 520-540, 1987.
[12] K. Barr and K. Asanovi´c. "Energy Aware Lossless Data Compression". In Proc. Of the ACM Conf. on Mobile Systems, Applications, and Services (MobiSys), May 2003.
[13] M. Barma, M. Riaz, R. Saidur and B. Long, "Estimatiion of Thermoelectric power generation by recovering waste heat from biomass fired thermal oil heater," Energy Conversion and Management, vol. 98, pp. 303-313, 2015.
[14] D. C. Burger, T. M. Austin, S. Bennett, \Evaluating Future Micro-processors {The Simplescaler Toolset," Tech. Rep. 1342, Univ. of Wisconsin, CS Dept., 1997.
[15] M.P. Tcheou, L. Lovisolo, M.V. Ribeiro, and E.A. Da Silva, M.A. Rodrigues, J.M. Romano, P.S. Diniz. "The compression of electric signal waveforms for smart grids: State of the art and future trends". IEEE Transactions on Smart Grid, vol. 5, no. 1, pp. 290-302, 2013..
[16] S. Dai, W. Liu, Z. Wang, and K. Li, P. Zhu, P. Wang. "An Efficient Lossless Compression Method for Periodic Signals Based on Adaptive Dictionary Predictive Coding". Applied Sciences. vol. 10, no. 14, pp.4918, Jan. 2020.
[17] Nkongolo, M. and Tokmak, M., 2023. Zero-Day Threats Detection for Critical Infrastructures. *arXiv e-prints*, pp.arXiv-2306.
[18] Nkongolo, M., 2023. Fuzzy feature selection with key-based cryptographic transformations. *arXiv preprint arXiv:2306.09583*.
[19] Nkongolo, M., 2023. Fuzzification-based Feature Selection for Enhanced Website Content Encryption. *arXiv preprint arXiv:2306.13548*.

## BIOGRAPHIES OF AUTHORS

| | |
|---|---|
| 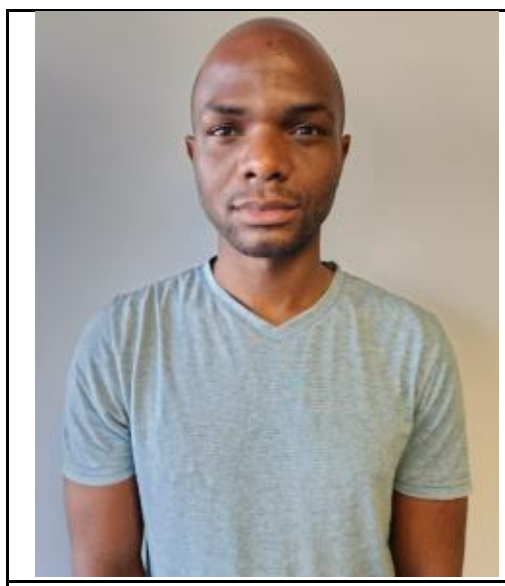 | **Tshimankinda Jerome Ngoy** earned his BEng degree in Electronics at University of Lubumbashi, in the Democratic Republic of the Congo (DRC), in 2011. He furthered his studies in South Africa, earning a BEng Honours in Microelectronics from University of Pretoria in 2018. He has supervised over 20 academia research projects in different levels, undergrad final year, B-Tech, honours, master thesis degrees (University of Pretoria, University of South Africa, University of Johannesburg, and Tshwane University of Technology). He has also authored of 3 conference papers. He is part-time lectuerer of Basic Analogue Electronics with the Faculty of Engineering, Independent Institute of Education (IIE) MSA, Johannesburg. He is currently work as an Electrical, Control and Instrumentation Engineer & Security System Engineer at Consulmet minerals, Johannesburg. He can be contacted at email: jerometshimak@gmail.com |
| | |
| 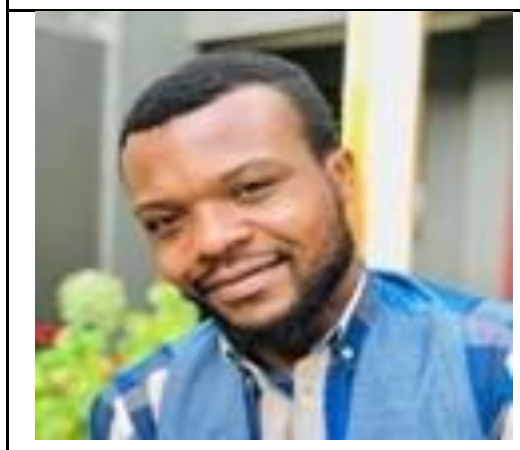 | **Wa Nkongolo Mike Nkongolo (PhD)** is a Lecturer in the Department of Informatics at the University of Pretoria. He holds a BSc in Information Technology (IT) from Universite Protestante de Lubumbashi in the Democratic Republic of the Congo (DRC). He furthered his studies in South Africa, earning a Hdip, BSc Honours, and master's degrees in computer science from the University of the Witwatersrand at the School of Computer Science and Applied Mathematics. He has recently completed his Ph.D. thesis in IT at the University of Pretoria in the department of Informatics. Prior to his academic career, Mike worked in the telecommunication industry as a Systems Engineer, Technical Support Specialist, and Data Analyst for NEC XON via Maven Systems Worx Support. He is actively involved in the academic community, serving as a reviewer for IEEE Transactions on Education, arXiv endorser (Cryptography, Information Retreival, and Machine Learning), and as a programme committee member (Computer Science Track) at the South African Computer Scientists & Information Technologists Conference. Mike's research interests span several areas, including Network Security (Cryptology, Deep Packet Inspection, Intrusion Detection/Prevention, Data Protection), Artificial Intelligence, Machine Learning, Information Retrieval, Natural Language Processing, and Game Theory. Currently, he is am member of the South African Council of Educator as well as the IITPSA (Institute of Information Technology Professionals South Africa), previously Computer Society of South Africa (CSSA). |
| | |